\documentclass[aps,prb, preprint, citeautoscript, showpacs,floatfix,superscriptaddress]{revtex4-2}
\usepackage{bm}
\usepackage{amssymb}
\usepackage{graphicx}
\usepackage{amsmath}
\usepackage{textcomp}
\usepackage{colordvi}
\usepackage{multirow}
\usepackage{ulem}
\usepackage{braket}
\usepackage{color}
\usepackage{booktabs}
\usepackage{xcolor}

\begin{document}
\title{Strain-Induced Tuning of Third-Harmonic Generation in Monolayer Black Phosphorene}

\author{Yan Meng}
%\email{D202410465@xs.ustb.edu.cn}
\affiliation{Department of Physics, Institute of Theoretical Physics, University of Science and Technology Beijing, Beijing 100083, China.} 
\author{Kainan Chang}
\email{knchang@ustb.edu.cn}
\affiliation{Department of Physics, Institute of Theoretical Physics, University of Science and Technology Beijing, Beijing 100083, China.} 
\author{Wei Song}
%\email{D202410465@xs.ustb.edu.cn}
\affiliation{Department of Physics, Institute of Theoretical Physics, University of Science and Technology Beijing, Beijing 100083, China.} 
%
%\author{Lingzhao Meng}
%\affiliation{School of Physics and Electronics, Hunan Normal University, Changsha, Hunan 410081, P. R. China.} 
%
%
\author{Yuwei Shan}
\email{yuweis@ciomp.ac.cn}
\affiliation{GPL Photonics Laboratory, State Key Laboratory of Luminescence Science and Technology, Changchun Institute of Optics, Fine Mechanics and Physics, Chinese Academy of Sciences, Changchun 130033, China.}
\author{Jin Luo Cheng}
%\email{jinluocheng.phys@gmail.com}
\affiliation{School of Physics and Laboratory of Zhongyuan Light, Zhengzhou University, Zhengzhou, China}
\author{Luxia Wang}
\email{luxiawang@sas.ustb.edu.cn}
\affiliation{Department of Physics, Institute of Theoretical Physics, University of Science and Technology Beijing, Beijing 100083, China.}

\begin{abstract}
Based on the tight-binding model and the semiconductor Bloch equations, this work systematically reveals the microscopic mechanism of strain engineering in turning of third-harmonic generation (THG) in monolayer black phosphorene (BP).
The results show that under strain-free conditions, monolayer BP exhibits significant in-plane anisotropy, and its dominant susceptibility component reaches a maximum of $\chi^{(3);xxxx} = 1.8 \times 10^{-17} \, \text{m}^2/\text{V}^2$, agreeing well with the experimental results.
By applying uniaxial and biaxial strains along the armchair ($x$), zigzag ($y$), and out-of-plane ($z$) directions, we find that the THG response presents strong direction dependence and unique spectral shifting behaviors: in-plane compressive strain and out-of-plane tensile strain both significantly enhance the THG conductivity and induce a redshift, whereas in-plane tensile strain and out-of-plane compression lead to suppression and a blueshift, with the tuning efficiency following the order of $z > y > x$. The microscopic origin of these phenomena is identified as the synergistic modulation of the bandgap and Berry connection by strain.
Furthermore, the synergistic or competitive effects of biaxial strain further enrich the manipulation of THG signals.
Strain engineering can serve as an effective strategy for dynamically controlling nonlinear optical processes in two-dimensional materials, and it also lays a theoretical foundation for the development of high-performance reconfigurable infrared photonic devices.
\end{abstract}
	
\maketitle
	
\section{Introduction}
\label{introduction}
Nonlinear optics (NLO) is a foundation of modern photonics, essential for imaging, light 
manipulation, photon generation, transmission, and detection\,\cite{Ahmed2021, Sirleto2023}. Its impact extends to 
telecommunications, information storage, computing, and biomedicine\,\cite{Ahmed2021, Sirleto2023}. Moreover, combining NLO with integrated optics, the nonlinear integrated photonic devices offer compact and efficient platforms for frequency conversion\,\cite{Dutt2024, Wang2024}.
Parametric processes---particularly 
second-harmonic generation (SHG) and third-harmonic generation (THG)---serve as critical probes for investigating the structural properties of micro- and nanomaterials due to their sensitivity to crystal structure and electronic properties. SHG and THG convert incident light at frequency \(\omega\) into output light at \(2\omega\) and \(3\omega\)\,\cite{Boyd2008}, respectively. 
Specifically, SHG originates from the second-order nonlinear susceptibility \(\chi^{(2)}\). 
Within the electric dipole approximation, this process adheres to strict symmetry, i.e., it is exclusively allowed in non-centrosymmetric crystals.
In contrast, THG, as a third-order nonlinear process 
governed by \(\chi^{(3)}\), is permitted in all materials.
Owing to its lack of symmetry restrictions, THG offers a versatile and robust optical method for determining crystallographic orientation and exploring anisotropic physical properties.
Furthermore, the frequency up-conversion inherent in NLO parametric processes enables the efficient conversion of optical responses from the near-infrared (NIR) spectral range to the visible domain. This capability not only overcomes the bandgap limitations of conventional detectors but also provides an innovative strategy for the development of broadband, polarization-sensitive
photodetectors and novel coherent light sources, holding significant promise for advancing all-optical signal processing and integrated photonics.	

The pursuit of efficient and tunable third-order nonlinear optical materials is pivotal for realizing high-performance integrated photonic devices. In recent years, two-dimensional (2D) materials have garnered significant attention due to their strong nonlinear optical susceptibility
\,\cite{Matos2023, Ahmed2021}. Among the vast family of 2D materials, black phosphorene (BP) stands out with 
the following advantages:
First, its highly anisotropic in-plane structure, arising from the puckered honeycomb lattice, results in drastically different optical responses along the armchair and zigzag directions. This intrinsic anisotropy provides an ideal platform for achieving polarization-dependent nonlinear optical functionalities.
Second, BP possesses a direct bandgap that is highly tunable with layer thickness, spanning a wide range from approximately 0.3 eV to 2 eV\,\cite{Yi2019}, which fills the gap between the zero bandgap of graphene and that of transition metal dichalcogenides (1.5–2.5 eV).
Furthermore, BP exhibits additional merits including high carrier mobility\,\cite{Ho2025}, excellent biocompatibility\,\cite{Sun2025}, and remarkable optical properties with multi-dimensional tunability under external fields (e.g., electric gating\,\cite{Chang2026}, magnetic field\,\cite{Yang2023}, strain\,\cite{Yarmohammadi2021}).
These attributes render BP a promising candidate for exploring nonlinear optical devices and developing next-generation anisotropic photonic devices. 

Experimentally, BP exhibits remarkable THG properties. Early studies employing near-infrared excitation confirmed the strong polarization anisotropy of the THG signal and revealed a unique thickness-dependent  THG response\,\cite{Youngblood2017}. Recent research has further extended the excitation wavelength from the near-infrared to the mid-infrared regime\,\cite{Zhu2025}, systematically unveiling the substantial THG potential of BP in this spectral range.
Despite these exceptional performances, the THG response of BP can be further dynamically modulated by external physical fields\,\cite{Zhang2021, Zhong2022, Huang2020}. 
Among these approaches, strain engineering is regarded as a ``universal key'' for tuning the electronic structure and optical properties of low-dimensional materials\,\cite{Katiyar2025}. Owing to their ultrathin nature and exceptional mechanical robustness, two-dimensional materials can sustain strains exceeding 10\% without fracture\,\cite{DiGiorgio2022}.
Strain engineering leverages the high strain sensitivity of two-dimensional materials to enable continuous tuning of the electronic band structure and associated physical properties, and this tuning strategy has been successfully realized in Nb$_2$SiTe$_4$\,\cite{Ouyang2025} and graphene\,\cite{Ji2025}.
From a technical standpoint, strain engineering in two-dimensional materials has been extensively validated through various experimental methodologies. Uniaxial tensile strain can be readily achieved using cantilever-based or four-point bending apparatus\,\cite{Suzuki2011a}. For biaxial compression, specialized electromechanical devices have been developed, including microheater actuators that induce biaxial expansion through thermal effects\,\cite{Ryu2020} and innovative setups capable of applying mechanical strain to two-dimensional materials at cryogenic temperatures while simultaneously probing the atomic lattice deformation via X-ray diffraction\,\cite{HenriquezGuerra2024}. These established experimental platforms confirm the practical feasibility of applying controlled uniaxial and biaxial strain to atomically thin materials, providing robust foundations for exploring strain-modulated nonlinear optical phenomena.
Using pressure\,\cite{Koenig2011}, concentrated forces\,\cite{Elibol2016}, etc. to load two-dimensional materials in the out-of-plane direction, this type of deformation is relatively controllable and can generate large strains. For example, in experiments, the strain of molybdenum disulfide (MoS$_2$) can be as high as 5.6\%\,\cite{Wang2015}.

For BP, theoretical calculations have long predicted its ability to withstand tensile strains up to 30\%\,\cite{Peng2014}, and to achieve continuous bandgap tuning---even semiconductor--semimetal transitions---through in-plane or out-of-plane strain, offering substantial flexibility for optoelectronic device design. 
However, the vast majority of current research remains focused on the influence of strain on the linear optical properties of BP\,\cite{Yarmohammadi2021}. In contrast, the potential of strain engineering to dynamically control third-harmonic generation (THG)—a nonlinear optical process critical for frequency conversion, all-optical switching, and high-resolution imaging—has remained largely unexplored.
This study aims to fill this critical gap by systematically exploring the tuning mechanisms of uniaxial and biaxial strains on the third-order nonlinear optical response of monolayer BP, as
schematically shown in Figs.\,\ref{band}\,(a) and (b).

This paper is organized as follows. In Sec.\,\ref{models}, we introduce the tight-binding model for monolayer BP and describe the treatment of strain effects via Harrison's rule. The third-order nonlinear conductivity for THG is derived using the semiconductor Bloch equations, and the symmetry of the conductivity tensor is analyzed. Section \,\ref{results} establishes the theoretical framework for strain-modulated THG, including the evolution of the band structure under uniaxial and biaxial strain, the strain dependence of THG spectra, and the analysis of Berry curvature contributions. The tuning efficiency along different strain directions and the synergistic effects of multi-dimensional strain are systematically investigated. We conclude in Sec.\,\ref{conclusions}. 

\section{THEORETICAL MODELS}
\label{models}
%	\begin{figure*}[htp]
%		\centering
%		\includegraphics[scale=0.1]{fig/bp-thg.png}
%		\includegraphics[scale=0.1]{fig/bp-sxsy.png}
%		\includegraphics[scale=0.1]{fig/bp-sz.png}
%		\caption{}
%		\label{bp}
%	\end{figure*}
	
	\subsection{Tight-Binding Method}
	
BP is a single-element layered crystal composed of phosphorene atoms arranged in a puckered orthorhombic lattice. 
The primitive basis vectors are 
$\bm{a} = a_0{\hat{\bm x}}$, $\bm{b} = b_0{\hat{\bm y}}$ and $\bm{c} = c_0{\hat{\bm z}}$, with lattice constants $a_0 = 4.43\,\text{\AA}$,  $b_0 = 3.27\,\text{\AA}$, and $c_0 = 5.46\,\text{\AA}$,
respectively\,\cite{Zhang2023, TaghizadehSisakht2015}. 
The unit cell of monolayer BP contains four atoms, with
atomic positions $\bm{\tau}_\alpha$ given by
$(- \frac{1}{2} + u) \bm a + \frac{1}{4} \bm b - v \bm c$,
$-  u \bm a + \frac{1}{4} \bm b + v \bm c$,
$u \bm a - \frac{1}{4} \bm b + v \bm c$,
$(\frac{1}{2} - u) \bm a - \frac{1}{4} \bm b - v \bm c$, where $u\approx0.16$, $v\approx0.20$  \cite{CastellanosGomez2014, Yuan2016, Zare2018, Le2018, Le2019a, Le2019b, Rudenko2015}. 
The low-energy electronic excitation of pristine monolayer BP is described by a four-band tight-binding model, and the corresponding Hamiltonian reads \cite{Rudenko2014, TaghizadehSisakht2015,Le2019, Pham2019, Hap2024, Le2019c, Yang2016, Zhang2023, Ezawa2014, Yarmohammadi2021}:
	\begin{align}
		H_{\bm k}^0=
		\begin{pmatrix}
			0 & A_{\bm{k}} & B_{\bm{k}} & C_{\bm{k}} \\
			A_{\bm{k}}^{*} & 0 & D_{\bm{k}} & B_{\bm{k}} \\
			B_{\bm{k}}^{*} & D_{\bm{k}}^{*} & 0 & A_{\bm{k}} \\
			C_{\bm{k}}^{*} & B_{\bm{k}}^{*} & A_{\bm{k}}^{*} & 0
		\end{pmatrix}\,,
	\end{align}
with matrix elements
	\begin{subequations}
		\begin{align}
			A_{k} &= t_{2}^{\parallel} + t_{5}^{\parallel}e^{-i\bm{k}\cdot\bm{a}}\,, \\
			B_{k} &= 4t_{4}^{\parallel}e^{-i(\bm{k}\cdot\bm{a} - \bm{k}\cdot\bm{b})/2}\cos(\bm{k}\cdot\bm{a}/2)\cos(\bm{k}\cdot\bm{b}/2)\,, \\
			C_{k} &= 2e^{i\bm{k}\cdot\bm{b}/2}\cos(\bm{k}\cdot\bm{b}/2)(t_{1}^{\parallel}e^{-i\bm{k}\cdot\bm{a}} + t_{3}^{\parallel})\,, \\
			D_{k} &= 2e^{i\bm{k}\cdot\bm{b}/2}\cos(\bm{k}\cdot\bm{b}/2)(t_{1}^{\parallel} + t_{3}^{\parallel}e^{-i\bm{k}\cdot\bm{a}})\,.
		\end{align}
	\end{subequations}
	Here the intralayer hopping energies are $t_{1}^{\parallel} = -1.220$ eV, $t_{2}^{\parallel} = 3.665$ eV, $t_{3}^{\parallel} = -0.205$ eV, $t_{4}^{\parallel} = -0.105$ eV, 
	and
	$t_{5}^{\parallel} = -0.055$ eV,
	 which are fitted from the first-principles calculation with GW corrections\,\cite{Rudenko2014, Yuan2015, Takao1981}.

	In crystalline materials, introducing strain can modify the electronic hopping parameters between atoms\,\cite{Yarmohammadi2021}.
	Let the initial position of an atom be the three-dimensional vector $\mathbf{r}_i$. When strain $\varepsilon_\gamma$ is applied along the $\gamma = x, y, z$ directions, the deformed position $\tilde{r}_i$ can be expressed as:
	\begin{align}
		\tilde{r}_i = \left( 1 + \sum_{\gamma = x,y,z} \alpha_i^\gamma \varepsilon_\gamma \right) r_i \,,
	\end{align}
	where $\alpha_i^\gamma$ are  dimensionless geometrical coefficients 
	(see Table \ref{tab:strain_coeff}).
	According to Harrison's rule\,\cite{Yarmohammadi2021}, the electronic hopping integral $t_i$ is inversely proportional to the square of the interatomic distance ($t_i \propto r_i^{-2}$), indicating that the hopping amplitude decreases with increasing atomic separation.
	Therefore, under strain, the modified hopping parameter $\tilde{t}_i$ can be approximated as:
	\begin{align}
		\tilde{t}_i \approx \left( 1 - 2 \sum_{\gamma} \alpha_i^\gamma \varepsilon_\gamma \right) t_i\, .
	\end{align}
	This shows that strain directly modulates the electronic hopping strength through the geometric factors $\alpha_i^\gamma$. 
	
	\begin{table}[h]
		\centering
		\caption{Strain coefficients $\alpha_i^\gamma$ for BP\,\cite{Yarmohammadi2021}.}
		\label{tab:strain_coeff}
		\begin{tabular}{lccccc}
			\toprule\toprule
			$\gamma$ & $\alpha_1^\gamma$ & $\alpha_2^\gamma$ & $\alpha_3^\gamma$ & $\alpha_4^\gamma$ & $\alpha_5^\gamma$ \\
			\midrule
			$x$ & 0.4460 & 0.0992 & 0.7505 & 0.3976 & 0.7530 \\
			$y$ & 0.5571 & 0      & 0.2461 & 0.2280 & 0      \\
			$z$ & 0      & 0.9052 & 0      & 0.3722 & 0.2538 \\
			\bottomrule\bottomrule
		\end{tabular}
	\end{table}	
	
	Diagonalization of the strain-modified Hamiltonian yields eigenenergies $\epsilon_{n\bm{k}}$ and eigenstates $C_{n\bm{k}}$ for band $n$ and wave vector $\bm k$, satisfying
	\begin{align}
		\hat{H}_{\bm{k}}C_{n\bm{k}} = \epsilon_{n\bm{k}}C_{n\bm{k}}.
	\end{align}
	The optical response
	requires the matrix elements of the position operator $\hat{\bm{r}}_{\bm{k}}$  and velocity operator $\hat{\bm{v}}_{\bm{k}}$ in the eigenbasis.
	In the Bloch representation, the position-operator matrix elements between atomic orbitals are
	\begin{align}	
		{\bm{r}}_{\alpha_1 \alpha_2\bm{k}} &= i \nabla_{\bm{k}} \delta_{\alpha_1 \alpha_2}+ \bm{\tau}_{\alpha_1} \delta_{\alpha_1 \alpha_2}\,,
	\end{align}
	where $\bm{\tau}_{\alpha_1}$ is the position of atom $\alpha_1$ 
	within the unit cell. 
	The  velocity operator is defined as
	\begin{align}	
		\hat{\bm{v}}_{\bm{k}} = \frac{1}{i\hbar} {[\hat{\bm{r}}_{\bm{k}}}, \hat{H}_{\bm{k}}]\,,
	\end{align}
	and its matrix elements between eigenstates are
	\begin{align}	
		\bm{v}_{n_1n_2\bm{k}} = C^\dagger_{n_1\bm{k}} \hat{\bm{v}}_{\bm{k}} C_{n_2\bm{k}}\,.
		\label{v}
	\end{align}
	Therefore, the Berry connection (defined by the off-diagonal position matrix elements) is given by
	\begin{align}
		\bm{\xi}_{n_1n_2\bm{k}} = C^\dagger_{n_1\bm{k}} \hat{\bm{r}}_{\bm{k}} C_{n_2\bm{k}}\,.
		\label{xi}
	\end{align}
	Because $\hat{\bm{r}}_{\bm{k}}$ contains a derivative with respect to $\bm{k}$, direct evaluation of Eq.\,(\ref{xi})  requires $\bm{k}$-smooth wavefunctions, which is numerically challenging due to phase arbitrariness. However, for $n_1 \neq n_2$, the off-diagonal elements can be obtained from the velocity matrix elements via \cite{Xiao2010, Sipe2000, Aversa1995, Hipolito2016, Chang2026}
	
	\begin{align}
		\bm{r}_{n_1n_2\bm{k}} = 
		\begin{cases} 
			\bm{\xi}_{n_1n_2\bm{k}} = \frac{\bm{v}_{n_1n_2\bm{k}}}{i\omega_{n_1n_2\bm{k}}} & \text{if } n_1 \neq n_2 \\
			0 & \text{if } n_1 = n_2
		\end{cases}\,,
		\label{r}
	\end{align}
	with $\hbar\omega_{n_1n_2\bm{k}} = \epsilon_{n_1\bm{k}} - \epsilon_{n_2\bm{k}}$. 	
\subsection{THG Conductivity}

We focus on the THG induced by a uniform optical field \(\bm{E}(t) = \bm{E}_0(t)e^{-i\omega t} + \text{c.c.}\) (where \(\bm{E}_0(t)\) is the slowly varying envelope function). In the studied low-dimensional system, the THG response is expected to be dominated by electronic mechanisms, as ionic contributions may be small\, \cite{Grillo2024}, unlike in bulk materials\, \cite{Prussel2023}. Correspondingly, using the electronic states described above, the electron dynamics under the applied optical field can be determined by solving the semiconductor Bloch equations in the length gauge\, \cite{Cheng2015, Aversa1995, Hipolito2016, Pedersen2015, Hipolito2018}. 
By applying perturbation theory with respect to the electric field (see details in our previous work\,\cite{Chang2026}), we obtain the third-order nonlinear optical conductivity and focus specifically on the THG component \(\sigma^{(3);dabc}(\omega, \omega, \omega)\).

Before proceeding to numerical calculations, we analyze the symmetry properties of the conductivity tensor. Monolayer BP has an orthorhombic crystal system\,\cite{Jiang2016}, leading to 21 independent nonzero elements in its third-order nonlinear
tensor\,\cite{Boyd2008}. For the THG conductivity tensor \(\sigma^{(3);dabc}\), nine of these are independent:
\begin{enumerate}
	\item[] \(xxxx\), \(xxyy=xyxy=xyyx\), \(xxzz=xzxz=xzzx\), 
	\item[] \(yyyy\), \(yyxx=yxyx=yxxy\), \(yyzz=yzyz=yzzy\), 
	\item[] \(zzzz\), \(zzxx=zxzx=zxxz\), \(zzyy=zyzy=zyyz\),
\end{enumerate}
where the intrinsic permutation symmetry \(\sigma^{(3);dabc}=\sigma^{(3);dbac}=\sigma^{(3);dcba}\) has been applied.
Since the contribution along the \(z\)-direction is relatively weak, for radiation normally incident on the phosphorene layer, only four independent nonzero components work: \(xxxx\), \(xxyy\), \(yyxx\), \(yyyy\).  
Therefore, the THG conductivity components under applied strain remain identical to those previously analyzed.
Besides, the effective third-order nonlinear susceptibility for THG is given by \cite{Chang2026}
\begin{align}
	\chi^{(3);dabc}_\text{eff}(3\omega) = -\frac{\sigma^{(3);dabc}(\omega, \omega,  \omega)}{3i\omega\epsilon_0 c_0}\,,
	\label{eff}
\end{align}
with the vacuum permittivity \(\epsilon_0\).

	\section{RESULTS AND DISCUSSIONS}
	\label{results}
	\subsection{Uniaxial Strain Effects on Band Structure}

	\begin{figure*}[htp]
			\centering
		    \hspace*{-0.5cm}\includegraphics[scale=0.1]{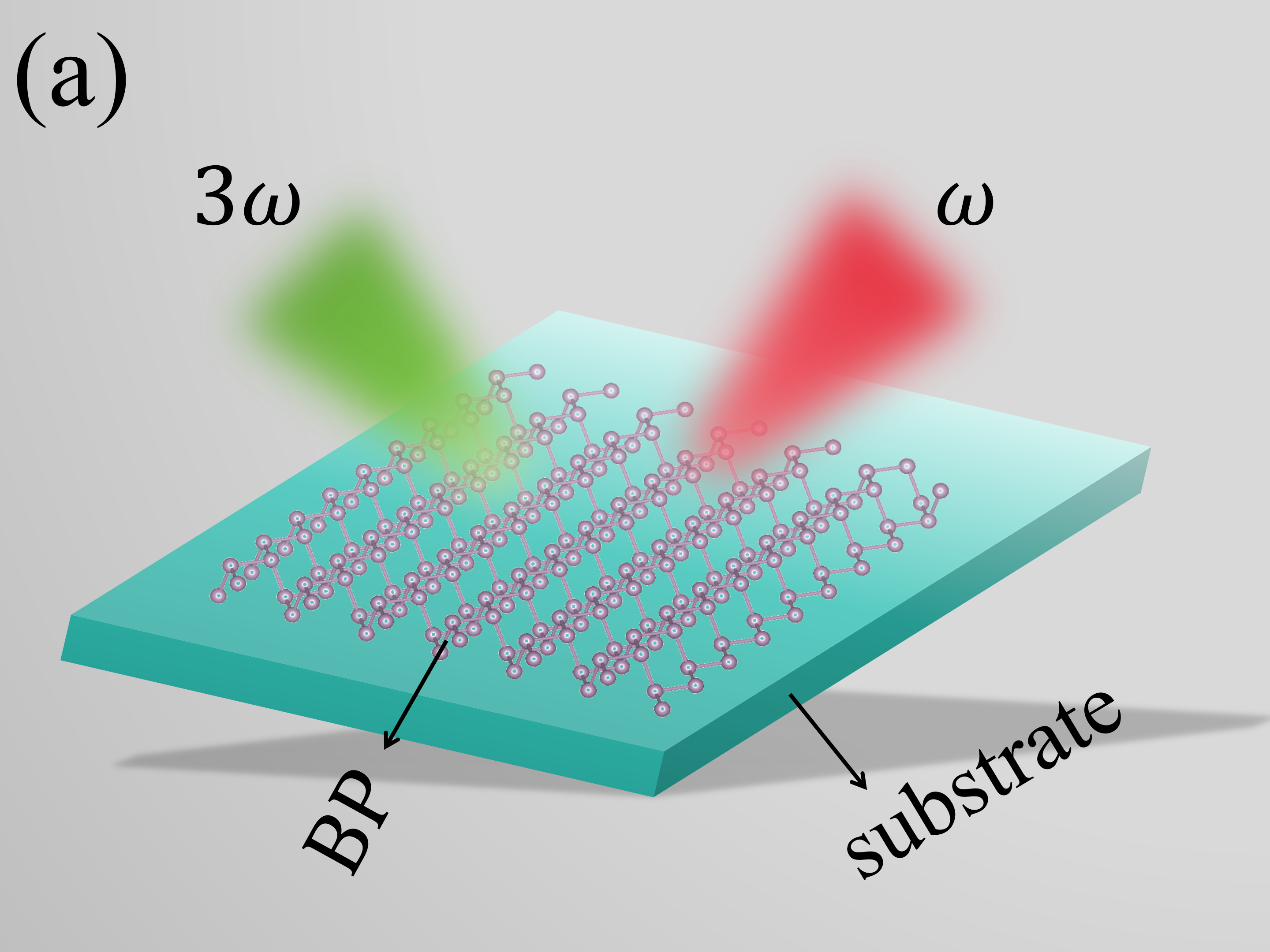}
			\hspace*{0.2cm}\includegraphics[scale=0.1]{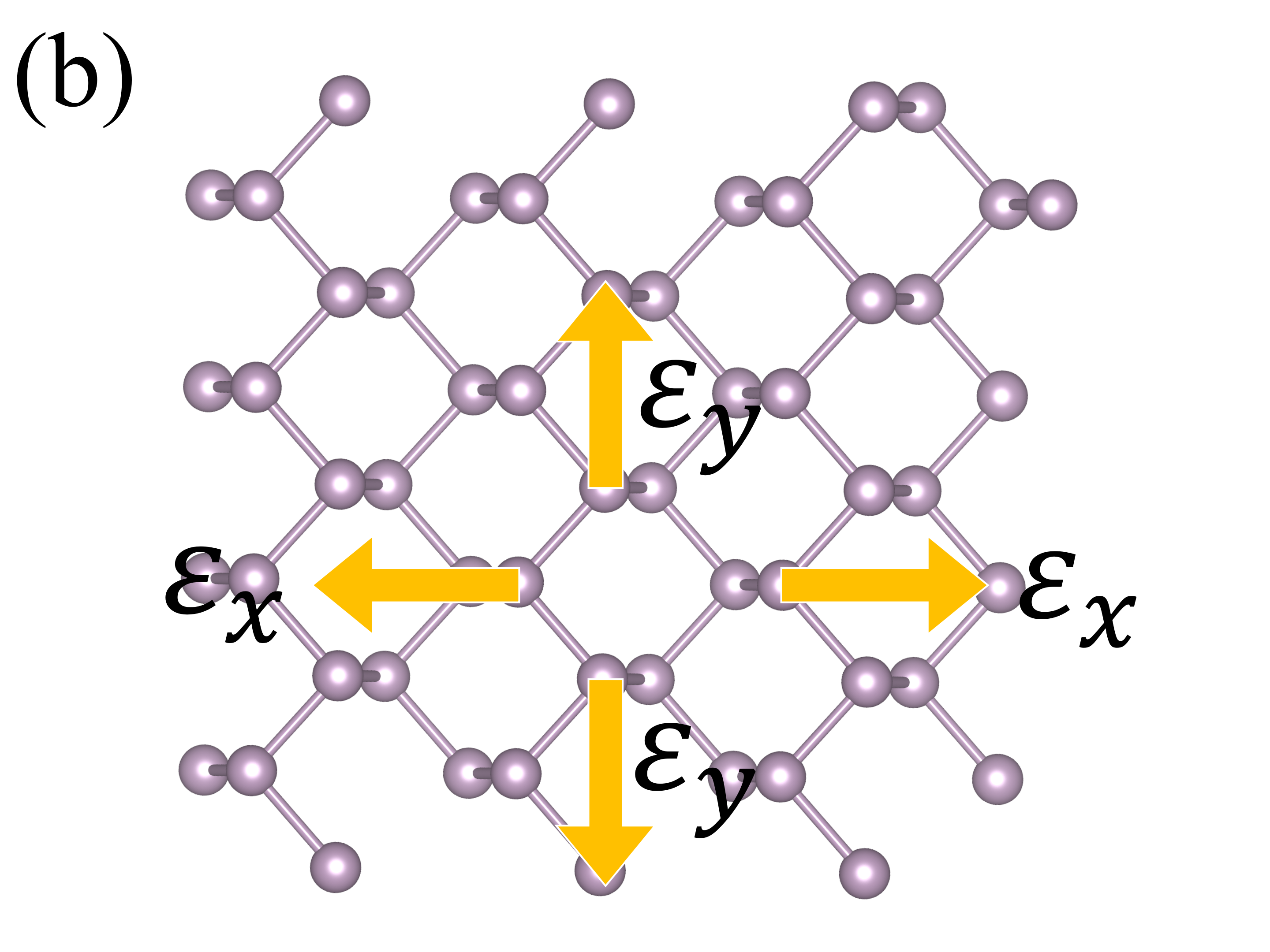}
			\includegraphics[scale=0.1]{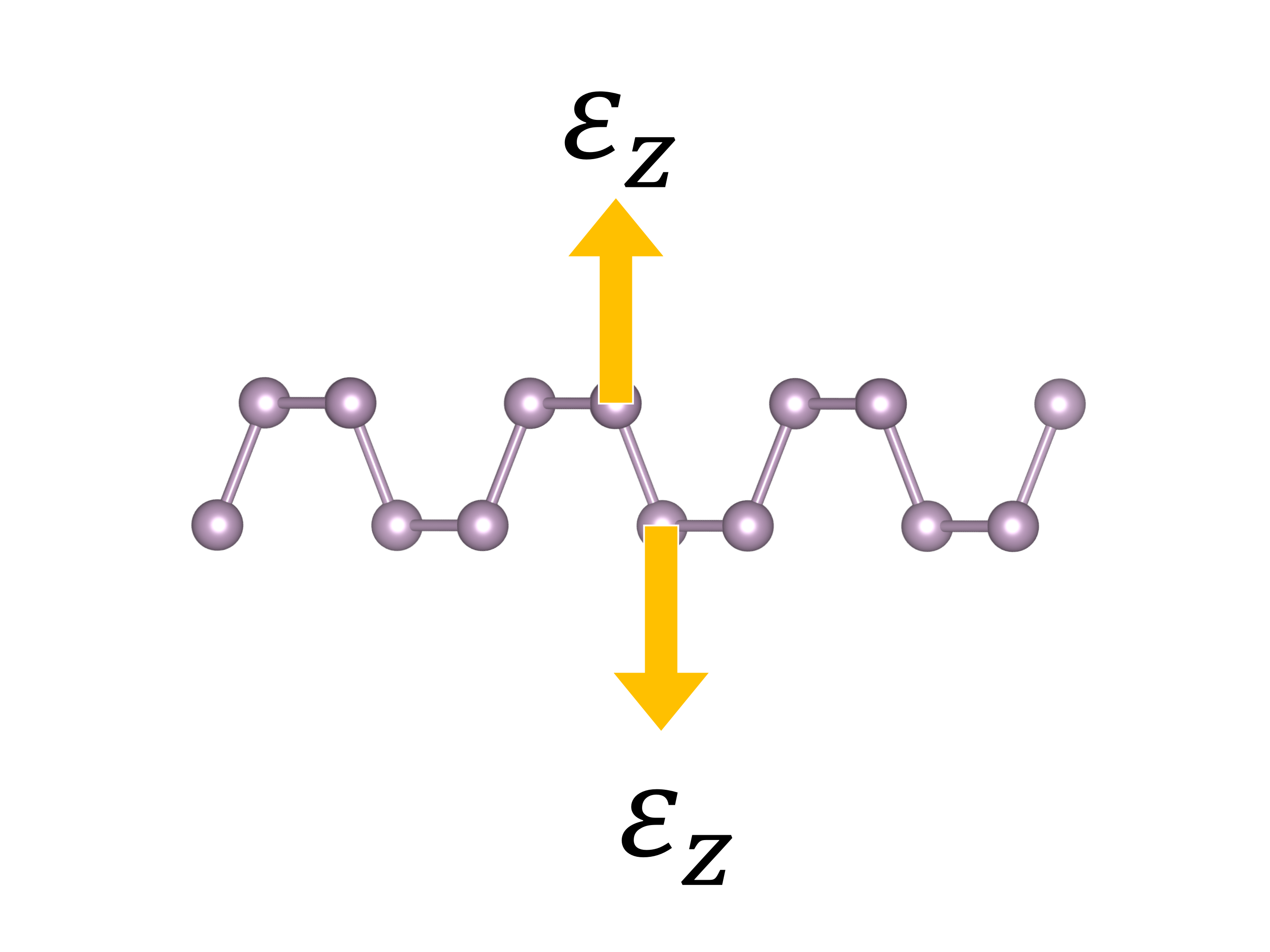}\\
			\includegraphics[scale=1]{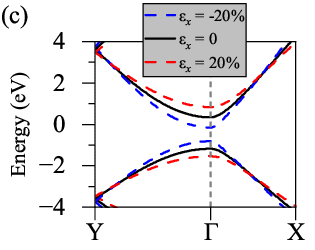}
			\includegraphics[scale=1]{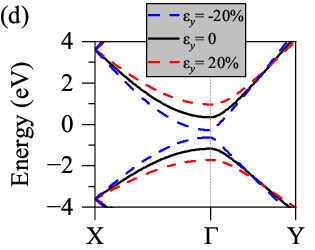}
		    \includegraphics[scale=1]{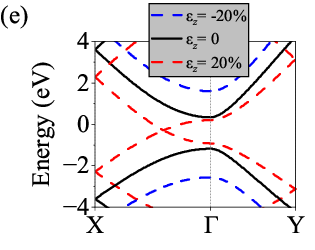}\\
		    \includegraphics[scale=1]{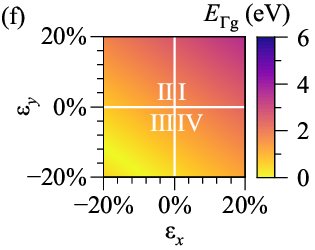}
		    \includegraphics[scale=1]{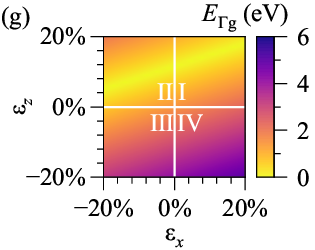}
		    \includegraphics[scale=1]{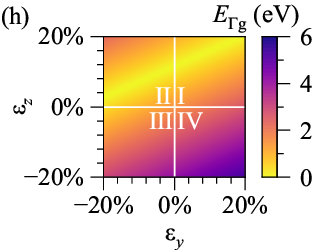}
    \caption{
	(a) Schematic illustration of THG in monolayer BP on a substrate, where incident light at frequency $\omega$ generates a nonlinear optical response at $3\omega$. 
	(b) Crystal structure of BP: top view showing in-plane uniaxial strains $\varepsilon_x$ (AC direction) and $\varepsilon_y$ (ZZ direction); side view illustrating out-of-plane strain $\varepsilon_z$. 
	(c--e) Evolution of the electronic band structure under uniaxial strain applied along (c) the $x-$, (d) $y-$, and (e) $z-$ directions.  
	(f--h) Contour plots of the energy difference $E_{\Gamma\text{g}}$ between the highest valence band and the lowest conduction band at the $\Gamma$ point under biaxial strain combinations, which divide the strain parameter space into four distinct regions (I--IV): (f) $\varepsilon_x$ versus $\varepsilon_y$, (g) $\varepsilon_x$ versus $\varepsilon_z$, and (h) $\varepsilon_y$ versus $\varepsilon_z$.}
	\label{band}
	\end{figure*}
	
We first analyze the influence of strain dimensionality on the band structure of monolayer BP. In our calculations, the strain parameter $\varepsilon_\gamma$ ranges from $-20\%$ to $+20\%$ to cover a broad strain range, ensuring a comprehensive understanding of the strain effect. 
Figures\,\ref{band}\,(c)-(e) illustrate the band structures under uniaxial strains along three distinct directions. 
Figures\,\ref{band}\,(f)-(h) show the energy difference (i.e., the direct bandgap) at the $\Gamma$ point between the highest valence band (index~2) and the lowest conduction band (index~3) under biaxial strain combinations, denoted as ${E}_{\Gamma\text{g}}=\epsilon_{3\Gamma}-\epsilon_{2\Gamma}$.
For the uniaxial in-plane strain, the bandgap exhibits a linear dependence, as shown in  Figs.\,\ref{band}\,(c) and (d). 
Under the strain-free condition $\varepsilon_x = \varepsilon_y = 0$, the original bandgap is $1.52$ eV.  When tensile strain is applied with $\varepsilon_x = +20\%$ (or $\varepsilon_y = +20\%$), the bandgap increases; conversely, under compressive strain with $\varepsilon_x = -20\%$ (or $\varepsilon_y = -20\%$), the bandgap decreases. In Fig.\,\ref{band}\,(f), 
setting $\varepsilon_x = 0$ corresponds to uniaxial strain purely along the $y$-direction; similarly, $\varepsilon_y = 0$ gives uniaxial strain purely along the $x$-direction.
The bandgap exhibits a monotonic dependence on strain, widening (narrowing) with increasing (decreasing) $\varepsilon_x$ or $\varepsilon_y$. Despite similar trends along both directions, strain applied in the $y$-direction induces a larger bandgap change, revealing a higher strain sensitivity. 
 
We then consider the effect of uniaxial out-of-plane
strain on the band structure. 
Figure\,\ref{band}\,(e) shows that the bandgap increases
under $\varepsilon_z = -20\%$, whereas the bands cross under $\varepsilon_z = 20\%$.
In contrast to the in-plane case, however,
continuously changing uniaxial out-of-plane strain, the bandgap evolves nonmonotonically, as shown in Fig.\,\ref{band}\,(g) and (h).
When $\varepsilon_x = \varepsilon_y =0$, as the tensile strain along the $z$-direction increases up to 20\%, the bandgap first decreases and then increases. Notably, at $\varepsilon_z \approx 11.5\%$, the bandgap reaches zero,
where the bands overlap and the system transitions to a semimetallic state. 
This nonmonotonic behavior featuring a zero bandgap constitutes a key evidence for the strain-induced semiconductor--semimetal--semiconductor phase transition along the $z$-direction. In contrast, compressive strain along the $z$-direction leads to a monotonic increase in the bandgap, and the system remains in a semiconducting state throughout. 

\subsection{Biaxial Strain Effects on Band Structure}
	
Building upon the study of uniaxial strain, we further investigate the synergistic control of the bandgap in monolayer BP under biaxial strain.
As shown in Figs.\,\ref{band}\,(f)--(h), dividing the biaxial strain parameter space into four quadrants systematically reveals the electronic phase transition behavior induced by different strain combinations.
Specifically, figure\,\ref{band}\,(f) shows that the in-plane
strain is simultaneously applied along the armchair ($x$) and zigzag ($y$) directions, the bandgap evolution exhibits clear regional dependence.
In quadrant~I ($\varepsilon_x > 0$, $\varepsilon_y > 0$), corresponding to biaxial tensile strain, the bandgap increases significantly with the increasing strain along both directions, gradually driving the system gradually from a semiconducting toward an insulating phase. 
In quadrant~III ($\varepsilon_x < 0$, $\varepsilon_y < 0$), under biaxial compression, the bandgap initially decreases continuously with increasing compressive strain until it vanishes, leading to a semiconductor--semimetal transition. 
Upon further compression, $E_{\Gamma\text{g}}$ slightly recovers, accompanied by band inversion.
In contrast, in quadrants~II and~IV (i.e., opposite-sign strain combinations, such as $\varepsilon_x > 0$ with $\varepsilon_y < 0$, or vice versa), the strain effects along the $x$‑ and $y$-directions largely cancel each other, resulting in only weak bandgap variations, and the system maintains stable semiconducting characteristics throughout. 

However, coupling in-plane and out-of-plane strains introduces a more complex competition/cooperation mechanism, whose outcome strongly depends on the sign combinations of the applied strains.
As shown in Figs.\,\ref{band}\,(g) and (h), in quadrant IV ($\varepsilon_{\text{in-plane}}>0$, $\varepsilon_{z}<0$), both in-plane tension and out-of-plane compression contribute to bandgap enlargement, 
producing a synergistic enhancement that keeps the system stably in a semiconducting phase or even pushes it toward an insulating phase. 
In quadrant~II ($\varepsilon_{\text{in-plane}} < 0,\ \varepsilon_z > 0$), by contrast, both in-plane compression and out-of-plane tension  tend to reduce the bandgap, causing it to drop sharply to zero and then band inversion.
In quadrants I and III, the two strains have opposite effects on the bandgap.
In quadrant~I ($\varepsilon_{\text{in-plane}} > 0$, $\varepsilon_z>0$), in-plane tensile strain tends to increase the bandgap, while out-of-plane tensile strain tends to reduce it; their competition leads to a slow decrease of the bandgap, which eventually closes to zero, driving the system into a semimetallic phase. 
In quadrant~III ($\varepsilon_{\text{in-plane}} < 0,\ \varepsilon_z < 0$), although in-plane compression tends to narrow the bandgap, this effect is dominated by the gain induced by out-of-plane compression, ultimately causing a slight increase in the bandgap, so the system remains semiconducting throughout this region.
In short, biaxial strain engineering offers a powerful tool for precisely tuning the electronic structure and quantum phases of monolayer BP, through systematic competition and synergy among strains applied along different directions.
	
\subsection{Effect of Uniaxial Strain on THG}

%\begin{figure*}[htp]
%	\centering
%	\includegraphics[scale=0.3]{fig/xxxx-sx.pdf}
%	\includegraphics[scale=0.3]{fig/xxyy-sx.pdf}\\
%	\includegraphics[scale=0.3]{fig/yyxx-sx.pdf}
%	\includegraphics[scale=0.3]{fig/yyyy-sx.pdf}
%	\caption{}
%	\label{sx}
%\end{figure*}
\begin{figure*}[htp]
	\centering
	\includegraphics[scale=1, trim=0 0.6cm 0 0, clip]{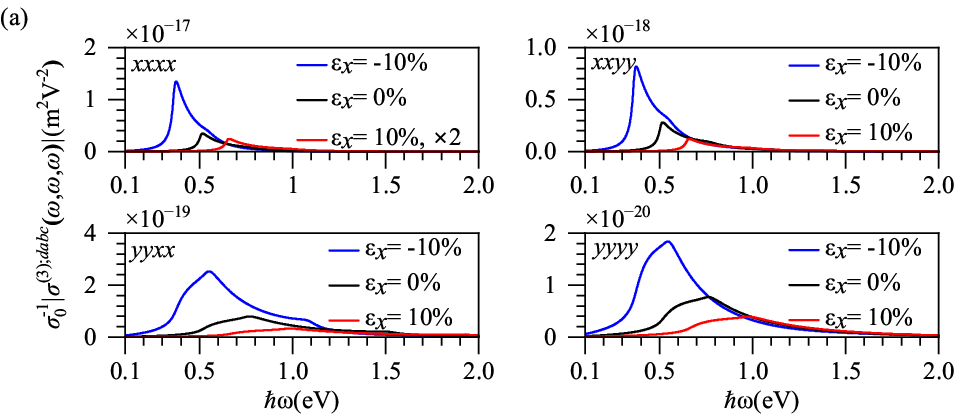}
	\includegraphics[scale=1, trim=0 0.6cm 0 0, clip]{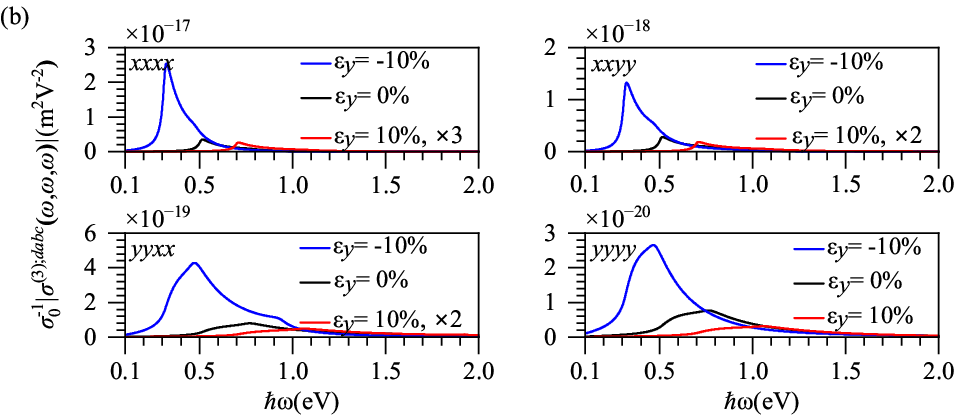}
	\includegraphics[scale=1]{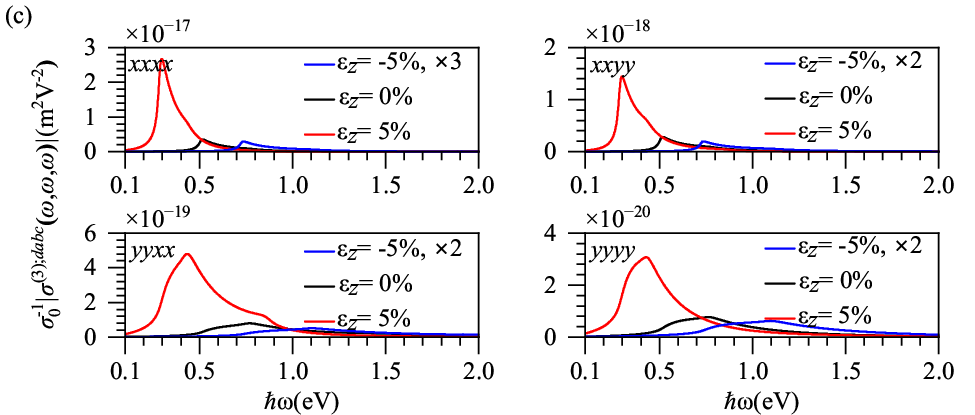}
	\caption{Spectra of THG conductivities ($\sigma^{(3);xxxx}$, $\sigma^{(3);xxyy}$, $\sigma^{(3);yyxx}$, and $\sigma^{(3);yyyy}$) of monolayer BP under uniaxial strains. The strains are applied along  (a)  armchair direction ($x$) with $\varepsilon_x = -10\%, 0\%, +10\%$, (b)  zigzag direction ($y$) with $\varepsilon_y = -10\%, 0\%, +10\%$, and (c) out-of-plane direction ($z$) with $\varepsilon_z = -5\%, 0\%, +5\%$. 
	Here $\sigma_0 = e^2/(4\hbar)$.}
	\label{thg}
\end{figure*}

\begin{figure*}[htp]
	\centering
	\includegraphics[scale=1]{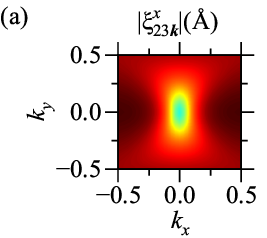}
	\includegraphics[scale=1]{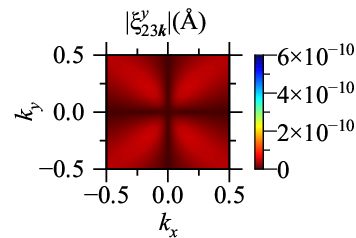}
	\includegraphics[scale=1]{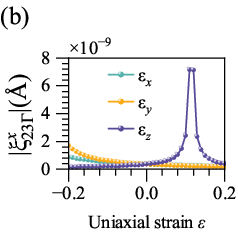}
	\caption{(a) Distribution of the absolute value of the Berry connection $|\xi_{23\bm{k}}|$ between the highest valence band and the lowest conduction band along the $x$ (left panel) and $y$ (right panel) directions in the 2D momentum space under strain-free conditions. (b) Evolution of the $x$-component (left panel) of the matrix element modulus $|\xi_{23\Gamma}|$ at the $\Gamma$ point (i.e., $\bm{k}=0$) as a function of uniaxial strain $\varepsilon$ applied along different directions.}  
	\label{berry}    
\end{figure*}
	
Next, we investigate the influence of strain on the THG response. Figure\,\ref{thg}(a) illustrates the dependence of the four principal components of the third-order nonlinear 
conductivities---
$\sigma^{(3);xxxx}$, $\sigma^{(3);xxyy}$, $\sigma^{(3);yyxx}$, 
and 
$\sigma^{(3);yyyy}$---on photon energy 
$\hbar\omega$
in monolayer BP under uniaxial strain applied along the 
armchair-direction ($x$-direction).
Under the strain-free condition
(corresponding to the bandgap
$E_{\Gamma\text{g}}=1.52$ eV), the THG spectra for  $\sigma^{(3);xxxx}$ and $\sigma^{(3);xxyy}$ have resonant
peaks at  $\hbar\omega=E_{\Gamma\text{g}}/3 = 0.52$ eV, 
whereas the resonant peaks for $\sigma^{(3);yyxx}$  and $\sigma^{(3);yyyy}$  appear at $\hbar\omega= E_{\Gamma\text{g}}/2 = 0.76$ eV. The former denotes a three-photon process, while the latter 
arises from a two-photon process.  
A comparison among the amplitudes of resonant peaks  reveals
the order
$|\sigma^{(3);xxxx}| > |\sigma^{(3);xxyy}| > |\sigma^{(3);yyxx}| > |\sigma^{(3);yyyy}|$, 
indicating significant optical anisotropy, with a notably stronger response along the in-plane armchair direction.
This anisotropy can be further traced to the contribution of the Berry connection between the valence and conduction bands. 
Specifically, $\sigma^{(3);xxxx}$ is primarily related to the Berry connection component $\xi^{x}_{23\bm{k}}$; $\sigma^{(3);yyyy}$ depends solely on $\xi^{y}_{23\bm{k}}$; while $\sigma^{(3);xxyy}$ and $\sigma^{(3);yyxx}$ depend on both. 
As shown in Fig.\,\ref{berry}(a), the amplitude of $|\xi^{x}_{23\bm{k}}|$ is significantly larger than that of $|\xi^{y}_{23\bm{k}}|$ overall, which is particularly prominent at the Brillouin zone center ($\Gamma$ point). This directly leads to the strongest response in $\sigma^{(3);xxxx}$ and the weakest in $\sigma^{(3);yyyy}$. This result shows the significant 
anisotropy on the monolayer BP's nonlinear optical response,
which is consistent with the linear absorption characteristics of monolayer BP\,\cite{Le2019}, and  
with our previous work on the second harmonic generation of 
few-layer BP\,\cite{Chang2026}.
To quantify this anisotropy, we calculate the THG susceptibility using Eq.\,(\ref{eff}).
At a resonant photon energy of $\hbar\omega = 0.52$ eV, $\chi^{(3);xxxx} = 1.8 \times 10^{-17}$ m$^2$/V$^2$ for BP, which agrees well with recent experimental results for BP ($\sim 10^{-17}$ m$^2$/V$^2$)\,\cite{Zhu2025}. This value is about two orders of magnitude larger than the theoretical value of undoped graphene ($3.25 \times 10^{-19}$ m$^2$/V$^2$ at $\hbar\omega = 0.72$ eV)\,\cite{Cheng2014}, and is close to the experimental measurement of graphene at a similar photon energy ($\sim 8 \times 10^{-17}$ m$^2$/V$^2$)\,\cite{Cheng2014}.

Uniaxial strain exerts a consistent and significant influence on the interband resonant peaks of each component of THG. 
For strain along the armchair direction as shown in Fig.\,\ref{thg}(a), compared with the strain-free case, compressive in-plane strain 
($\varepsilon_x = -10\%$) induces a red shift of the resonant peak, directly related to the reduction in bandgap; while tensile strain along the same direction ($\varepsilon_x = +10\%$) leads to a blue shift, consistent with the increase in bandgap. This indicates that the influence of strain on the nonlinear optical response is closely coupled with its effect on the band structure. In addition, compressive strain markedly enhances the peak intensity, whereas tensile strain weakens it. 
When the strain is applied along the $y$-direction, as shown in Fig.\,\ref{thg}\,(b), the trend of interband resonance peak shifts remains consistent with that induced by $x$-direction strain. 
However, the $y$-direction strain produces a more pronounced change in intensity, indicating a higher nonlinear optical sensitivity to strain along the $y$-direction.
When the uniaxial strain is applied along the out-of-plane ($z$-) direction, as shown in Fig.\ref{thg}(c), these four main THG tensor components exhibit a distinct trend in the THG peak position and intensity: out-of-plane compression causes a blue shift accompanied by a clear reduction in intensity, whereas out-of-plane tension induces a red shift and substantially enhances the nonlinear response. 
Such a behavior is exactly opposite to that under in-plane $x$- and $y$-direction strains.
This distinct physical mechanism can be attributed to the distinct evolution of the interband Berry connection at the $\Gamma$ point under in-plane versus out-of-plane strains. 
Since the $y$-component of the Berry connection is negligibly small overall, here we only need to focus on the variation of the $x$-component with strain in Fig.\,\ref{berry}(c), which exhibits a monotonic trend.
In-plane tensile (compressive) strain significantly suppresses (enhances) the Berry connection. In contrast, out-of-plane strain induces an opposite trend, reaching a maximum where the bandgap closes, which demonstrates a strong directional dependence. Notably, the evolution of the Berry connection is strongly correlated with the bandgap magnitude: a narrower bandgap corresponds to a larger Berry connection. This relationship provides a clear physical explanation for the enhanced THG intensity accompanying the redshift of the THG peak.
Remarkably, an out-of-plane strain of only $\pm5\%$ achieves a THG variation comparable to that induced by a $\pm10\%$ in-plane strain along the $y$-direction, and even surpasses the tuning efficiency of a $\pm10\%$ in-plane strain along the $x$-direction. This further highlights a strong directional dependence in tuning efficiency, following the order $z > y > x$. The varying rates of the Berry connection under different strain directions, as illustrated in Fig.\,\ref{berry}, 
provide a clear physical explanation for this phenomenon.

\begin{figure*}[htp]
	\centering
	\includegraphics[scale=1]{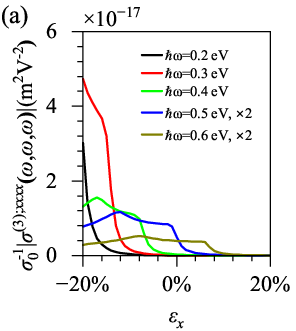}
	\includegraphics[scale=1]{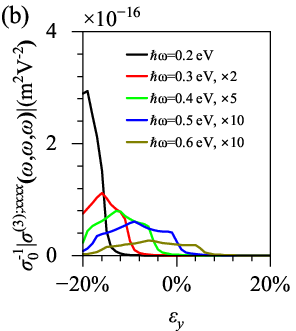}
	\includegraphics[scale=1]{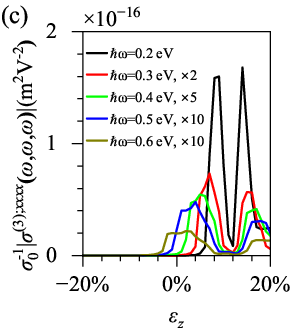}
	\caption{The $\sigma^{(3);xxxx}$ component of the third-order nonlinear conductivity in monolayer BP as a function of uniaxial strain along (a) $x$- (b) $y$- (c) and $z$-direction, for different incident photon energies.}
	\label{xxxx-hw}
\end{figure*}

Having established the photon-energy dependence of the THG response, we now examine how uniaxial strain tunes the THG intensity at fixed photon energies $\hbar\omega=0.2$, 0.3, $\cdots$, and 0.6 eV.
As shown in the Figs.\ref{xxxx-hw}(a) and (b), under in-plane compressive strain conditions ($\varepsilon_x < 0$ or $\varepsilon_y < 0$), the THG response at all photon energies is significantly enhanced.
As the photon energy decreases, a larger in-plane compressive strain is required for the THG conductivity to reach its maximum value, and the maximum value itself increases significantly. 
In contrast, under in-plane tensile strain ($\varepsilon_x > 0$ or $\varepsilon_y > 0$), the THG response is suppressed.
On the other hand, the mechanism of out-of-plane strain is more intricate. 
Figure\,\ref{xxxx-hw}(c) shows that, under out-of-plane compression ($\varepsilon_z < 0$), the THG response is very weak, while under out-of-plane tension ($\varepsilon_z > 0$), the response is markedly enhanced and exhibits two distinct maxima at specific strain values. This behavior originates from the strain-induced band-inversion process under out-of-plane tension. 
Note, although the band-inversion occurs for $\varepsilon_z > 11.5\%$, the THG response behaves similarly to that of a semiconductor, because the effective optical transition are dominated $\Gamma$ point. Therefore, we refer to this state as a semiconducting state. This means that the stain  can induce a  semiconductor--semimetal--semiconductor triple phase-transition pathway, which has already been demonstrated in the linear response regime\,\cite{Yarmohammadi2021}.

\subsection{Effect of Biaxial Strain on THG}
%\begin{figure*}[htp]
%	\centering
%	\includegraphics[scale=0.2]{fig/xxxx-sxsy.pdf}
%	\includegraphics[scale=0.2]{fig/xxxx-sxsz.pdf}
%	\includegraphics[scale=0.2]{fig/xxxx-sysz.pdf}
%	\caption{}
%	\label{}
%\end{figure*}

\begin{figure*}[htp]
	\centering
	\includegraphics[scale=1]{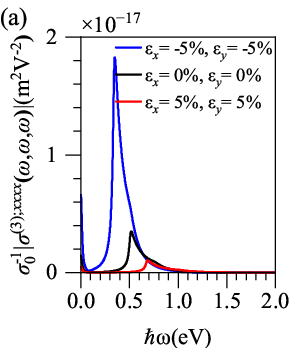}
	\includegraphics[scale=1]{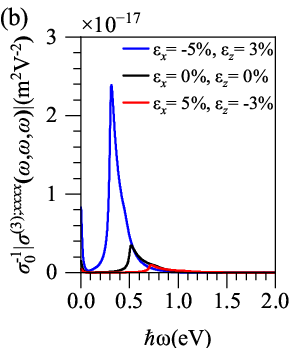}
	\includegraphics[scale=1]{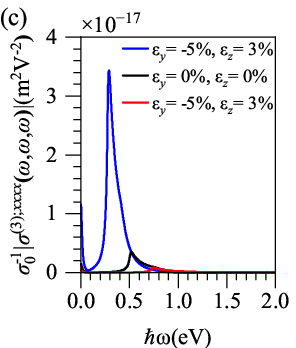}
	\caption{
		The $\sigma^{(3);xxxx}$ component of the third-order nonlinear conductivity of monolayer BP under various biaxial strain configurations:  
		(a) coupled in-plane $x$ and in-plane $y$ strains;  
		(b) coupled in-plane $x$ and out-of-plane $z$ strains;  
		(c) coupled in-plane $y$ and out-of-plane $z$ strains.
	}
	\label{s-two}
\end{figure*}

Finally, simultaneously applying strains along two different directions, we consider three typical biaxial strain configurations: (1) $\varepsilon_x = \varepsilon_y = \pm 5\%$ (biaxial in-plane strain);
(2) $\varepsilon_x=\pm5$ \%, $\varepsilon_z=\mp3$ \%; and (3) $\varepsilon_y=\pm5$ \%, $\varepsilon_z=\mp3$ \% (coupled in-plane and out-of-plane strain). All calculations are performed for the $\sigma^{(3);xxxx}$ tensor component.
Figure\,\ref{s-two}(a) shows that biaxial in-plane compression (simultaneous compression in both $x$ and $y$) enlarges the bandgap, which not only red-shifts the resonance peak but also significantly enhances the THG intensity. Conversely, biaxial in-plane tension (simultaneous tension in $x$ and $y$) reduces the bandgap, resulting in a blue shift of the THG peak and a pronounced weakening of the response. 
The spectra in Figs.\ref{s-two}(b) and (c) further reveal that both in-plane compression (tension) and out-of-plane tension (compression) tend to increase (decrease) the bandgap. Specifically, when applied synergistically (e.g., $x/y$ compression combined with $z$ tension), the bandgap increases substantially, leading to a pronounced red shift of the THG resonance peak and a strong enhancement of the nonlinear response. On the other hand, the combination of in-plane tension and out-of-plane compression (e.g., $x/y$ tension plus $z$ compression) cooperatively reduces the bandgap, blue-shifts the THG peak, and markedly suppresses the response. 

\section{CONCLUSIONS}
In this work, by combining the tight-binding model with the semiconductor Bloch equations, we systematically investigate the strain-induced THG modulation mechanism in monolayer BP.
The THG response exhibits strong intrinsic anisotropy.
In the unstrained case, the anisotropy of THG follows the order $|\sigma^{(3);xxxx}| > |\sigma^{(3);xxyy}| > |\sigma^{(3);yyxx}| > |\sigma^{(3);yyyy}|$. 
It is attributed to the fact that the $x-$component of the Berry connection between the valence and conduction bands is obviously larger than its $y-$component.
The effective susceptibility reaches a maximum of
$\chi^{(3)} = 1.8\times10^{-17}~\text{m}^2/\text{V}^2$ 
at resonant energy of $\hbar=0.52$ eV, which agrees well with experiments and is two orders of magnitude larger than that of graphene. Under uniaxial strain, in-plane compression induces a redshift and enhances the THG intensity, while in-plane tension causes a blueshift and suppresses the intensity. Out-of-plane compression leads to a blueshift and reduction of the THG intensity, whereas out-of-plane tension produces a redshift and a significant enhancement, with the tuning efficiency following the order $z > y > x$. Low-energy photons exhibit the highest strain sensitivity. Under biaxial strain, the synergistic or competitive effects of different strain components on the bandgap enable diversified control of the THG signal.

The above strain-induced modulation of the THG peak originates from the strain engineering of the band structure and the Berry connection. In-plane uniaxial strain tunes the direct bandgap (tension increases the gap, compression decreases it). Out-of-plane strain induces more complex behavior: tension can first close and then reopen the bandgap (at $\Gamma$ point).
In contrast to uniaxial strain, biaxial strain exhibits a superposition of strain effects from multiple directions, which markedly enhances the control over strain-induced band inversion and facilitates the occurrence of the above phenomenon.
realizing a semiconductor--semimetal--semiconductor phase transition, while compression only monotonically increases the bandgap.  
The redshift (blueshift) of the THG peak under strain corresponds to a decrease (increase) of the bandgap, while the enhancement (suppression) of the peak intensity originates from the increase (decrease) of the Berry connection as the bandgap narrows (widens), establishing a negative correlation.

This work establishes strain engineering as a powerful and versatile tool for dynamically tuning higher-order nonlinear optical processes in monolayer BP.
The unique anisotropic response and strain tunability of BP make it an ideal candidate for reconfigurable nonlinear photonic devices, including tunable frequency converters, anisotropic modulators, and on-chip integrated light sources operating in the near-infrared to mid-infrared spectral range.
\label{conclusions}

\bibliography{thgref}
\end{document}